# An EFFECTIVE WEB DOCUMENT CLUSTERING for INFORMATION RETRIEVAL

Rajendra Kumar Roul[1] (rkroul@bits-goa.ac.in), Dr.S.K.Sahay[2] (ssahay@bits-goa.ac.in)
*BITS, Pilani - K.K. Birla, Goa Campus, Zuarinagar, Goa - 403726, India.*

*Abstract*-The size of web has increased exponentially over the past few years with thousands of documents related to a subject available to the user. With this much amount of information available, it is not possible to take the full advantage of the World Wide Web without having a proper framework to search through the available data. This requisite organization can be done in many ways. In this paper we introduce a combine approach to cluster the web pages which first finds the frequent sets and then clusters the documents. These frequent sets are generated by using Frequent Pattern growth technique. Then by applying Fuzzy C-Means algorithm on it, we found clusters having documents which are highly related and have similar features. We used Gensim package to implement our approach because of its simplicity and robust nature. We have compared our results with the combine approach of (Frequent Pattern growth, K-means) and (Frequent Pattern growth, Cosine_Similarity). Experimental results show that our approach is more efficient then the above two combine approach and can handles more efficiently the serious limitation of traditional Fuzzy C-Means algorithm, which is sensitive to initial centroid and the number of clusters to be formed.

*Keywords*-Clustering, Fuzzy C-Means, Frequent Pattern growth, Gensim, Vector space model

## 1. INTRODUCTION

There is a tremendous growth of information on web. As the number of users using the web growing rapidly, so it creates many challenges of information retrieval which become the current research topics. In general the search results returned using any searching paradigm are not clustered automatically. But as the case is documents returned for a keyword may be of different nature depending upon the different meanings of the keyword. That is to say that the set of documents returned for a given keyword may further be subdivided into subsets of documents conveying similar sense of the keyword. Clustering the set of results will do this further sub-division and will present the results in a better way. It organizes the documents in such a way that the documents belonging to a group (cluster) are more similar to each other than to the ones which are a part of a different subgroup. Web mining has fuzzy characteristics, so fuzzy clustering is sometimes better suitable in comparison with conventional clustering. There are two basic methods of fuzzy clustering [5], one which is based on fuzzy c-partitions, is called a Fuzzy C-Means(FCM) clustering and another based on the fuzzy equivalence relations, is called a Fuzzy Equivalence Clustering. Data mining technique called association analysis, which is useful for discovering interesting relationship hidden in large data set also useful for clustering. There are two broad principles use for association analysis [1].One is Apriori and another is Frequent Pattern(FP) growth principle. FP-growth is a divide and conquer strategy that mines a complete set of frequent itemsets without candidate generation. FP-growth outperformance Apriori because Apriori incurs considerable I/O overhead since it requires making several passes over the transaction data set. In this paper a method is being proposed of web document clustering based on FP-growth and FCM that helps the search engine to retrieve relevant web documents needed for any user. Documents in the FCM are strongly correlated; however traditional FCM clusters are sensitive to the initialization of membership matrix and center. It also needs the number of clusters to be formed as initial parameter. Our approach handles all this by using FP-growth approach which initializes this for FCM.

The paper is organized on the following lines: Section 2 covers the related work based on different clustering techniques used for web document. Section 3 describes the materials and methods used in approach. In section 4, we describe proposed approach adopted to form the clusters. The results are covered in section 5 and finally conclusion is presented in section 6.

## 2. RELATED WORK

The rapid growth of the web increases the web pages index in search engine. There are currently two major approaches taken to improve the search engine results. One approach is personalized web search and another is categorizing or grouping results into group that are meaningful to the user. Group of similar results form a cluster, which is an unsupervised attempt to find documents that are similar to each other and group them together. Many researchers are working in this area to find better clusters of documents. A.K.Jain et al[2] provides an extensive survey of various clustering techniques. Nicholas et al[3] presented the recent development in document clustering. Oren Zamir et al[4] in their research listed the key requirements of web document clustering methods as relevance, browsable summaries, overlap, snippet tolerance, speed and accuracy. They have given STC(Suffix Tree Clustering) algorithms which creates clusters based on phrased shared between documents. Shen huang et al[6] projected that web document clustering purpose a novel feature co-selection, which is called multitype feature co-selection for clustering(MFCC). MFCC uses intermediate clustering results in one type space to help selection in another type of feature space. Sun Park and others[7] proposed the document clustering methods using weighted semantic features and cluster similarities is done by using NMF(non negative matrix factorization). Srinivas et al[8] discussed a clustering algorithm using Incremental hierarchical clustering algorithm. Maufo Liy et al[9] purposed a web fuzzy clustering model. In their paper the experimental result of web fuzzy clustering in web user clustering proves the feasibility of web fuzzy clustering in web usage mining. Michael Steinbach et al[10] presented the result of an experimental study of some common documents clustering algorithms. Our approach used web documents clustering based on FP-growth and FCM that can help the user to find most relevant documents from the huge web and can able to handle the limitations of existing FCM. We use Gensim package[12] to avoid the dependency of the large training corpus size, and its ease of implementing vector space model.

## 3. MATERIALS AND METHODS

### 3.1 Vector Space Model

In vector space model, each document is defined as a multidimensional vector of keywords in Euclidean space whose axis correspond to the keyword i.e., each dimension corresponds to a separate keyword [11]. The keywords are extracted from the document and weight associated with each keyword determines the importance of the keyword in the document. Thus, a document is represented as, $D_j = (w_{1j}, w_{2j}, w_{3j}, w_{4j},...............,w_{nj})$ where, $w_{ij}$ is the weight of term i in document j indicating the relevance and importance of the keyword words.

*3.1.1 TF-IDF*

TF is the measure of how often a word appears in a document and IDF is the measure of the rarity of a word within the search index. Combining TF-IDF[11] is used to measure the statistical strength of the given word in reference to the query. Mathematically, $TF_i = n_i/(\Sigma_k n_k)$ where, $n_i$ is the number of occurrences of the considered terms and $n_k$ is the number of occurrences of all terms in the given document. $IDF_i = (\log N)/df_i$ where, N is the number of occurrences of the considered terms and $df_i$ is the number of documents that contain term i. $TF\text{-}IDF = TF_i \times IDF_i$

*3.1.2 Cosine_ Similarity Measure*

It is a technique to measure the similarity[11] between the document and the query. The angle($\Theta$)between the document and the query vector determines the similarity between the document and the query and it is written as

$$\cos \Theta = (\Sigma w_{q,j} w_{i,j})/(\sqrt{\Sigma w^2_{q,j}} \sqrt{\Sigma w^2_{i,j}}) \quad \text{(Eq. 1)}$$

where, $\sqrt{\Sigma w^2_{q,j}}$ and $\sqrt{\Sigma w^2_{i,j}}$ are the length of the query and document vector respectively.

If $\Theta = 0^0$, then the document and query vector are similar. As $\Theta$ changes from $0^0$ to $90^0$, the similarity between the document and query decreases i.e. document (D2) will be more similar to query than document (D1), if the angle between D2 and query is smaller than the angle between D1 and query.

### 3.2 Gensim

Gensim package is a python library for vector space modeling, aims to process raw, unstructured digital texts ("plain text"). It can automatically extract semantic topics from documents, used basically for the Natural Language Processing (NLP) community. Its memory (RAM) independent feature with respect to the corpus size allows to process large web based corpora.

### 3.3 FP-growth Algorithm

*3.3.1 Algorithm 1* (FP-tree construction)
Input: A transaction database *D* and a minimum support threshold *ξ*.
Output: FP-tree, the frequent-pattern tree of *D*.
Method: The FP-tree is constructed as follows.
1. Collect the set of frequent items($F_{items}$) and their support counts after scanning the transaction database(D) once. Sort $F_{items}$ according to descending support count as $L_{freq}$ the list of frequent items.

2. Create the root of an FP-tree, and label it as "null". For each transaction $I_{trans}$ in D do the following,

Select and sort the frequent items in $I_{trans}$ according to the order of $L_{freq}$. Let the sorted frequent list in $I_{trans}$ be [*e* | $E_{list}$], where *e* is the first element and $E_{list}$ is the remaining list. Call *insert_tree*([*e* | $E_{list}$],*T*), which is performed as follows.

Procedure *insert_tree*([*e* | *E*$_{list}$],*T*)

if *T* has a child *N* such that *N*.item-name=*e*.item-name, then increment *N*'s count by 1; else create a new node *N*, and let its count be 1, its parent link be linked to *T*, and its node-link to the nodes with the same item-name via the node-link structure. If *E*$_{list}$ is nonempty, call *insert_tree*(*E*$_{list}$,*N*) recursively.

*3.3.2 Algorithm 2* (FP-growth*: Mining frequent patterns with FP-tree by pattern fragment growth*)
Input: A database *D*, represented by FP-tree constructed according to Algorithm 1, and a minimum support threshold $\xi$ .
Output: The complete set of frequent patterns.
Method: *call FP-growth*(FP-tree*, null*).
Procedure *FP-growth*(*Tree, α*)
{
(1) *if Tree* contains a single prefix path
(2) *then* {
(3) *let P* be the single prefix-path part of *Tree*;
(4) *let Q* be the multipath part with the top branching node replaced by a *null* root;
(5) *for each* combination (denoted as *β*) of the nodes in the path *P do*
(6) *generate* pattern *β* ∪ *α* with *support = minimum support of nodes in β*;
(7) *let freq_ pattern_ set*(*P*) be the set of patterns so generated; *}*
(8) *else let Q* be *Tree*;
(9) *for each* item *a$_i$* in *Q do {*
(10) *generate* pattern *β = a$_i$* ∪ *α* with *support = a$_i$. .support*;
(11) *construct β*'s conditional pattern-base and then *β*'s conditional FP-tree *Tree$_β$* ;
(12) *if Tree$_β$* ≠ Φ
(13) *then call FP-growth*(*Tree$_β$, β*);
(14) *let freq_ pattern_ set*(*Q*) be the set of patterns so generated; *}*
(15) *return*(*freq_ pattern_ set*(*P*) ∪ *freq_ pattern_ set*(*Q*) ∪ (*freq_ pattern_ set*(*P*)×*freq_ pattern_ set*(*Q*)))
*}*

### 3.4 Fuzzy C-Means Algorithm

Fuzzy clustering allows each feature vector to belong to more than one cluster with different membership degrees (between 0 and 1) and vague or fuzzy boundaries between clusters. In the present discussion, the optimal number of clusters is same as the number of frequent item sets obtained using FP-growth.

FCM clustering is based on optimizing the following objective function. We try to get the minimum possible value of this function.

$$J_m = \sum_{i=1}^{N} \sum_{j=1}^{C} u_{ij}^m \|x_i - c_j\|^2$$

where,

m ϵ [1, ∞] - Fuzzy Coefficient

$u_{ij}$ - membership degree of $x_j$ w.r.t to $c_j$; range [0, 1]

$c_j$ - centroid(vector) of cluster j;

C - number of clusters

N - number of data vectors

$x_i$ - Data vector

FCM starts with random initial matrix or membership matrix U and, a fixed number of clusters. Number of columns and rows of the matrix U depends on the documents and the number of clusters.

Initially we have the cluster centres $c_j$ 's. Using the following updating formula, we iteratively find the valus of $u_{ij}$ and $c_j$ i.e clusters center vector updated with each iteration.

$$u_{ij} = \frac{1}{\sum_{k=1}^{C}(\|x_i - c_j\|/\|x_i - c_k\|)^{2/(m-1)}}$$

$$c_j = \frac{\sum_{i=1}^{N} u_{ij}^m \cdot x_i}{\sum_{i=1}^{N} u_{ij}^m}$$

with all symbols having same meaning as before and 'k' is the iteration step. The membership values are calculated w.r.t to the new centers. Belongingness of the document to the cluster is calculated using Euclidian distance between the center and the data point.The iterative process will stops when $\|U^{(k+1)} - U^{(k)}\| < \varepsilon$ where, $\varepsilon < 1$ is the termination criterion. This will converge to a local minimum of the function $J_m$ . The value of membership matrix U at that point will give the final cluster membership of the functions where C will give the centre points of the clusters.

### 3.5 Finding Initial Cluster Centers

For finding number of clusters and initial cluster centers, FP-growth algorithm for finding frequent item-sets has been used.

The frequent sets generated are of frequency greater than the minimum support supplied by the user. In the generated frequent sets of documents, the terms are taken to be the transactions and the documents are the items of the transactions. In this way the frequent sets generated are the ones which have particular set of terms in common and hence are closely related. This helps by deciding the number of clusters and also the centers of these clusters which is simply the centroid of the respective frequent item-set.

## 4. PROPOSED APPROACH

The algorithm used for clustering the web documents is described in this section and is represented graphically in Fig. 1.

Input:
1. Document set, $D$, to be clustered.
2. Value of minimum support, min_sup, to be used in FP-growth.
3. Value of fuzziness parameter, $m$.

Output:
1. Membership matrix, $U$, which shows how much a document belongs to a cluster
2. Matrix containing centers of the clusters, $C$.

Steps:

1. *Preprocessing:*
Preprocess the $D$ as follows:
• Remove the stop and unwanted words.
• Select noun as the keywords from $D$ and ignore other categories, such as verbs, adjectives, adverbs and pronounce.
• Do stemming using porter algorithm [13].
• Save each processed $n$ pages of $D$ as document $D_k$, where k = 1, 2, 3,…, n.

2. *Create term document matrix:*

Term document matrix, $T$, is created by counting the number of occurrences of each term in each document $D_k$. Each row $t_i$ of $T$ shows a term's occurrence in each document $D_k$.

3. *Extraction of frequent sets:*

FP-growth algorithm is used to extract maximal frequent sets of documents from the term document matrix $T$ using the value of minimum support( min_sup), given as an input and stored in $F$.

4. *Document vectors:*

Compute TF and IDF score for all the keywords of each $D_k$ and make document vectors of all the retrieved pages.

5. *Calculation of initial cluster centroids:*

Initial cluster centroids, $c_i$, are calculated using the maximal frequent sets obtained in Step 3. For each frequent set $f_i$ present in $F$, cluster $c_i$ is calculates as
$c_i = (D_1 + D_2 + … + D_j)/j$
where $j$ is the number of documents in frequent set $f_i$.
$C = \{ c_i : c_i$ is the centroid vector for cluster i $\}$.

6. *Calculation of final clusters:*

Final clusters are calculated using the cluster centroids $C$, and the fuzziness parameter $m$ and applying FCM algorithm on the set of document vectors, $D_k$ and the membership matrix $U$, and final cluster centroids $C$ are obtained where each $u_i$ gives the belongingness of each document vector in $D_k$ to the cluster $i$.

*Flow Diagram*: The following figure shows the steps to obtain the cluster of web documents which has discussed in the proposed approach.

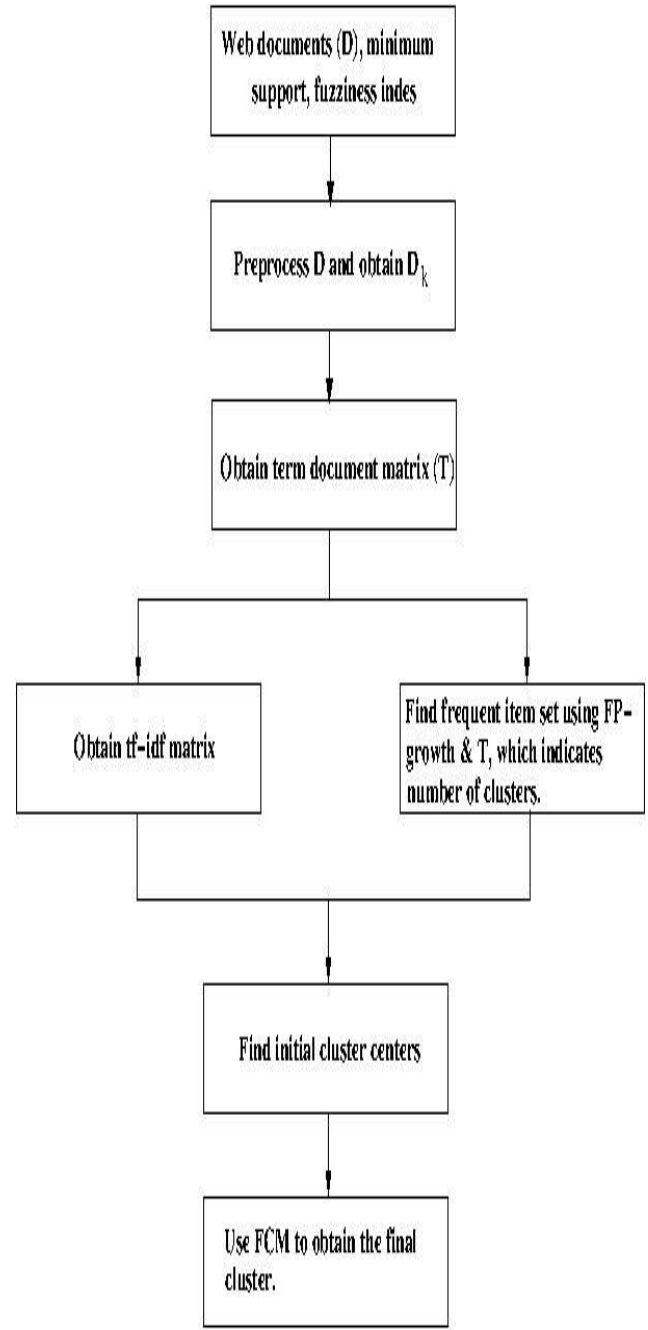

Figure 1: Web document clustering.

# 5. EXPERIMENTS AND RESULTS

This section provides the descriptions and characteristics of the data sets used for performing our experiments. Also, we briefly review the K-means and cosine_similarity techniques for comparison with our approach. The discussions on performance evaluation presented at the end.

## 5.1 Experimental data sets

A sample of 10 documents, given in Table 1, has been taken to explain the approach. These documents broadly categories to two topics namely Social network (D1 to D6) and Computer network (D7 to D10).The documents after preprocessing and extracting nouns as keywords are shown in Table 2. Keywords and their respective id's shown in table 3. Table 4 shows the term document matrix. The minimum support we have taken here is *3*. The FP-tree based on FP-growth algorithm derived from table 4 shown in Figure 2. Table 5 shows the tf-idf values for each keyword of each document. Frequent item-sets generated using FP-growth with the given value of minimum support are shown in Table 6. We have consider these two frequent sets shown in table 6 as two initial clusters and their initial centers are calculated shown in Table 7.Further these initial centers given as input to fuzzy c-means algorithm which generates the final clusters. Initial V matrix = [C1; C2] i.e. V is a 2x30 matrix with row 1 as C1 and row 2 as C2. The membership matrix is shown in Table 8. For both values of m=2 1nd m=1.5, the clusters are not going to be change. The final centroids are given in Table 9. Table 10 shows the final clusters.

## 5.2 K-means Technique

It consists the following steps:

1. Take the data space, which needs to be clusters.
2. Pre-determine the number of clusters as k.
3. Initialize k- means for the data, viz, m1, ….mk.
4. Using Euclidean distance, find the distance of data points from the mean.
5. Group the data points having minimum distance to the mean, to the corresponding mean.
6. Calculate the new mean of the each group formed in step 5.
7. Repeat steps 4-6 until the new mean formed is same as the previous mean.

## 5.3 Cosine_similarity Technique

1. Find out the initial clusters, $C_i$, using steps 1-3 of proposed approach.
2. Find out the centers of each $C_i$ using steps 4 and 5 of proposed approach.
3. Check for the uncluster documents, $UD_k$, from the document set $D$ mentioned in proposed approach.
4. Find the similarities of $UD_k$ to every $C_i$'s centers using Eq. 1.
5. Assign the document to that cluster which has maximum similarity. In case similarities between two or more clusters are same for any document, then assign the document to anyone these clusters.
6. Repeat the steps 4 and 5 till all $UD_k$ assigned to their respective $C_i$.

## 5.4 Discussion on Performance Evaluation

We have used the following metrics namely, Entropy and Purity for evaluating the performance of the proposed approach. The evaluation metrics are given below,

entropy = $\sum_{i=1}^{K}(m_i/m)E_i$, where K is the number of clusters and 'm' is the total number of documents. $m_j$ is the number of documents in cluster i.

$E_i = -\sum_{j=1}^{C}(p_{ij} \log_2 p_{ij})$, where C is the number of classes. $p_{ij} = m_{ij}/m_i$, where $m_{ij}$ is the number of documents of cluster i belongs to class j.

purity = $\sum_{i=1}^{K}(m_i/m)p_i$, where $p_i = \max_j (p_{ij})$ and all terms having same meaning as above.

We have compared our results with both K-means and cosine_similarity based clustering techniques, where both have been combine with FP-growth algorithm and the same experimental data sets has been used with minimum support *3*. The FP-growth algorithm used to give the number of frequent item sets which in turn gives the number of clusters to be form for K-means and each set center taken as the initial centroid for K-means. This concepts also used for cosine_similarity, where each document which is more similar to the center of a cluster will be added to that cluster by using Eq. 1. The details are listed in appendix. After comparison, we found the following results.

Final Clusters using FP-growth + Cosine_Similarity are {D1, D2, D3, D4, D6}, {D5, D7, D8, D9, D10}

Final Clusters using FP-growth+ K-means are {D1, D2, D3, D4, D6}, {D5, D7, D8, D9, D10}

Final Clusters using FP-growth + FCM are {D1, D2, D3, D4, D5, D6}, {D7, D8, D9, D10}

FP-growth+ Cosine_Similaritity

| Clusters | Social N/w Class | Computer N/W Class | Total | Entropy | Purity |
|---|---|---|---|---|---|
| Cluster 1 {D1,D2,D3,D4,D6} | 5 | 0 | 5 | 0 | 1 |
| Cluster 2 {D5,D7,D8,D9,D10} | 1 | 4 | 5 | .7 | .8 |
| Total | 6 | 4 | 10 | .35 | .9 |

FP-growth+ K-means

| Clusters | Social N/w Class | Computer N/W Class | Total | Entropy | Purity |
|---|---|---|---|---|---|
| Cluster 1 {D1,D2,D3,D4,D6} | 5 | 0 | 5 | 0 | 1 |
| Cluster 2 {D5,D7,D8,D9,D10} | 1 | 4 | 5 | .7 | .8 |
| Total | 6 | 4 | 10 | .35 | .9 |

FP-growth + FCM

| Clusters | Social N/w Class | Computer N/W Class | Total | Entropy | Purity |
|---|---|---|---|---|---|
| Cluster 1 {D1,D2,D3,D4,D5,D6} | 6 | 0 | 6 | 0 | 1 |
| Cluster 2 {D7,D8,D9,D10} | 0 | 4 | 4 | 0 | 1 |
| Total | 6 | 4 | 10 | 0 | 1 |

Overall Comparison

| Algorithm | Entropy | Purity |
|---|---|---|
| FP-growth + Cosine_similarity | .35 | .9 |
| FP-growth+ K-means | .35 | .9 |
| FP-growth + Fuzzy_C-means | 0 | 1 |

## 6. CONCLUSION

The proposed approach used FP-growth and FCM algorithms for clustering the web documents. This approach keeps the related documents in the same cluster so that searching of documents becomes more efficient. Experimental results show that our approach of FP-tree combine with FCM gives better results in terms of entropy and purity compare to traditional K-means and cosine_similarity techniques for clustering the web documents. Also it can handle the limitations of existing FCM. Future work would focus on improving the cluster sets by semantic based clustering and ranking the documents in each cluster using topic based modeling.

**ACKNOWLEDGMENT**

We are thankful to Bharat Deshpande and our colleagues Aruna Govada and K.V. Santhilata for their useful discussions and valuable suggestions.

**REFERENCES**

1. Introduction to Data Mining by Tan, Kumar and Steinbach, Pearson Education, 2006.

2. A.K.Jain, M.N.Murty and P.J Flynn, "Data clustering: A review," ACM computing surveys, 31(3):264-323, Sept 1999.

3. Nicholas O. Andrews and Edward A. fox, "Recent Development in Document Clustering Techniques",Dept of Computer Science, Virgina Tech 2007.

4. O.Zamir and O.Etzioni,"Web document clustering: A feasibility demonstration",in Proceeding of 19th International ACM SIGIR Conference on Research and Development in Information Retrieval, June 1998.

5. Klir & Yuan, "Fuzzy sets and Fuzzy Logic: Theory and Applications", Prentice Hall Publication.

6.Shen huang, Zheng chen, Yong yu and wei-ying ma,"Multitype Features Coselection for Web Document Clustering", IEEE transactions on knowledge and data engineering, Vol. 18, no 4,April 2006.

7.Sun Park, Dong Un An, Choi Im Cheon, "Document Clustering Method using Weighted Semantic Features and Cluster Similarity," digitel, pp.185-187,2010 Third IEEE International Conference on Digital Game and Intelligent Toy Enhanced Learning,2010.

8. Srinivas and C. Krishna Mohan, Efficient Clustering Approach using Incremental and
Hierarchical Clustering Methods. IEEE, 2010.

9. Maofu Liu, Yanxiang He and Huijun Hu," Web Fuzzy Clustering and Its Applications in Web Usage Mining", Proceedings of 8th International Symposium on Feature Software Technology (ISFST-2004).

10. Michael Stinbach, George Karypis and Vipin Kumar, "A Comparison of Document Clustering Techniques", KDD Workshop on Textmining, 2000

11. http://www.miislita.com/term-vector/term-vector-3.html
12.http://www.nlp.fi.muni.cz/projekty/gensim/intro.html
13.http://tartarus.org/martin/PorterStemmer/def.txt

# APPENDIX

| D_Id | Document |
|---|---|
| D1 | A dedicated website or other application that enables people to communicate with each other by posting information |
| D2 | Uses special sites to allow people to create a profile and form communities based on common interests |
| D3 | A social network service is an online service, platform, or site that focuses on building and reflecting of social networks or social relations among people |
| D4 | Social networks can be thought of as communities based upon interest or commonality that use the Internet to connect the people of the network |
| D5 | A group of people who exchange information and experience for professional or social purposes |
| D6 | Networking is establishing an informal communities of contacts among people with common social and business interests as a source of prospects, for the exchange of information, and for support |
| D7 | A computer network is a group of computers and devices interconnected by communications channels that facilitate communications among users and allows users to share resources and data |
| D8 | Computer Networking is the joining of two or more computers in order for them to communicate or jointly access a server. |
| D9 | A group of two or more computers linked by cables or wireless signals or both, which can communicate with one another using network protocols |
| D10 | A group of computers together with the sub-network or inter-network through which they can exchange data is called a computer network |

Table 1: Sample documents.

| D_Id | Document Keywords |
|---|---|
| D1 | website application people information |
| D2 | website people profile community interest |
| D3 | network service service platform website network relation people |
| D4 | network community interest commonality internet people network |
| D5 | group people information experience purpose |
| D6 | networks contact community people interest prospect information support |
| D7 | computer network group computer device channel communication user user resource data |
| D8 | computer network computer server |
| D9 | group computer cable signal network protocol |
| D10 | group computer network network data computer network |

Table 2: Documents after preprocessing.

| Keyword | Token Id. |
|---|---|
| website | 0 |
| application | 1 |
| people | 2 |
| information | 3 |
| profile | 4 |
| communities | 5 |
| interests | 6 |
| network | 7 |
| service | 8 |
| platform | 9 |
| relation | 10 |
| commonality | 11 |
| internet | 12 |
| group | 13 |
| contact | 14 |
| experience | 15 |
| purpose | 16 |
| prospects | 17 |
| support | 18 |
| computer | 19 |
| device | 20 |
| channel | 21 |
| communication | 22 |
| user | 23 |
| resource | 24 |
| data | 25 |
| server | 26 |
| cable | 27 |
| signal | 28 |
| protocol | 29 |

Table 3: Keywords and respective token ID's

| Keywords | D1 | D2 | D3 | D4 | D5 | D6 | D7 | D8 | D9 | D10 |
|---|---|---|---|---|---|---|---|---|---|---|
| website | 1 | 1 | 1 | 0 | 0 | 0 | 0 | 0 | 0 | 0 |
| application | 1 | 0 | 0 | 0 | 0 | 0 | 0 | 0 | 0 | 0 |
| people | 1 | 1 | 1 | 1 | 1 | 1 | 0 | 0 | 0 | 0 |
| information | 1 | 0 | 0 | 0 | 1 | 1 | 0 | 0 | 0 | 0 |
| profile | 0 | 1 | 0 | 0 | 0 | 0 | 0 | 0 | 0 | 0 |
| community | 0 | 1 | 0 | 1 | 0 | 0 | 0 | 0 | 0 | 0 |
| interest | 0 | 1 | 0 | 1 | 0 | 1 | 0 | 0 | 0 | 0 |
| network | 0 | 0 | 2 | 2 | 0 | 1 | 1 | 1 | 1 | 3 |
| service | 0 | 0 | 2 | 0 | 0 | 0 | 0 | 0 | 0 | 0 |
| platform | 0 | 0 | 1 | 0 | 0 | 0 | 0 | 0 | 0 | 0 |
| relation | 0 | 0 | 1 | 0 | 0 | 0 | 0 | 0 | 0 | 0 |
| commonality | 0 | 0 | 0 | 1 | 0 | 0 | 0 | 0 | 0 | 0 |
| internet | 0 | 0 | 0 | 1 | 0 | 0 | 0 | 0 | 0 | 0 |
| group | 0 | 0 | 0 | 0 | 1 | 0 | 1 | 0 | 1 | 0 |
| contact | 0 | 0 | 0 | 0 | 0 | 1 | 0 | 0 | 0 | 0 |
| experience | 0 | 0 | 0 | 0 | 1 | 0 | 0 | 0 | 0 | 0 |
| purpose | 0 | 0 | 0 | 0 | 1 | 0 | 0 | 0 | 0 | 0 |
| prospect | 0 | 0 | 0 | 0 | 0 | 1 | 0 | 0 | 0 | 0 |
| support | 0 | 0 | 0 | 0 | 0 | 1 | 0 | 0 | 0 | 0 |
| computer | 0 | 0 | 0 | 0 | 0 | 0 | 1 | 2 | 1 | 2 |
| device | 0 | 0 | 0 | 0 | 0 | 1 | 0 | 0 | 0 | 0 |
| channel | 0 | 0 | 0 | 0 | 0 | 0 | 1 | 0 | 0 | 0 |
| communication | 0 | 0 | 0 | 0 | 0 | 0 | 1 | 0 | 0 | 0 |
| user | 0 | 0 | 0 | 0 | 0 | 0 | 2 | 0 | 0 | 0 |
| resource | 0 | 0 | 0 | 0 | 0 | 1 | 0 | 0 | 0 | 0 |
| data | 0 | 0 | 0 | 0 | 0 | 0 | 1 | 0 | 0 | 1 |
| server | 0 | 0 | 0 | 0 | 0 | 0 | 0 | 1 | 0 | 0 |
| cable | 0 | 0 | 0 | 0 | 0 | 0 | 0 | 0 | 1 | 0 |
| signal | 0 | 0 | 0 | 0 | 0 | 0 | 0 | 0 | 1 | 0 |
| protocol | 0 | 0 | 0 | 0 | 0 | 0 | 0 | 0 | 1 | 0 |

Table 4: Term-document matrix.

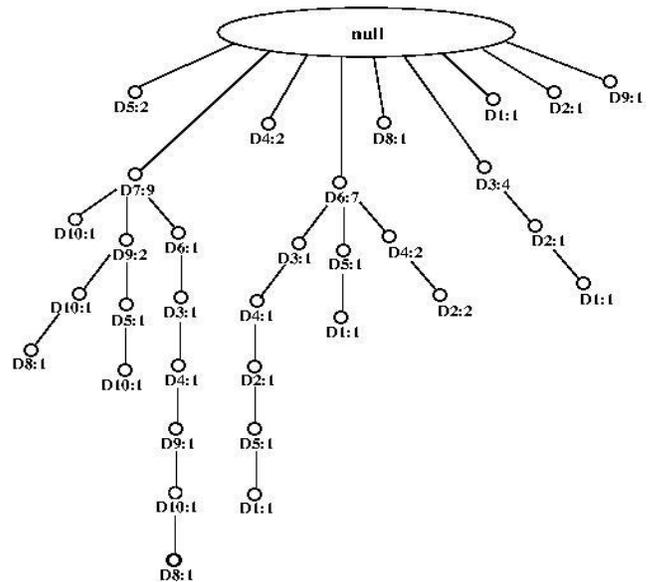

Figure 2: FP-tree representation of the web document.

|          | TF * IDF |       |       |       |       |       |       |       |       |       |
|----------|-------|-------|-------|-------|-------|-------|-------|-------|-------|-------|
| Keywords | D1    | D2    | D3    | D4    | D5    | D6    | D7    | D8    | D9    | D10   |
| website  | 0.131 | 0.105 | 0.065 | 0     | 0     | 0     | 0     | 0     | 0     | 0     |
| application | 0.25 | 0   | 0     | 0     | 0     | 0     | 0     | 0     | 0     | 0     |
| people   | 0.055 | 0.044 | 0.028 | 0.032 | 0.044 | 0.028 | 0     | 0     | 0     | 0     |
| information | 0.131 | 0  | 0     | 0     | 0.105 | 0.065 | 0     | 0     | 0     | 0     |
| profile  | 0     | 0.2   | 0     | 0     | 0     | 0     | 0     | 0     | 0     | 0     |
| community | 0    | 0.105 | 0     | 0.075 | 0     | 0.065 | 0     | 0     | 0     | 0     |
| interest | 0     | 0.105 | 0     | 0.075 | 0     | 0.065 | 0     | 0     | 0     | 0     |
| network  | 0     | 0     | 0.039 | 0.044 | 0     | 0.019 | 0.014 | 0.039 | 0.026 | 0.066 |
| service  | 0     | 0     | 0.25  | 0     | 0     | 0     | 0     | 0     | 0     | 0     |
| platform | 0     | 0     | 0.125 | 0     | 0     | 0     | 0     | 0     | 0     | 0     |
| relation | 0     | 0     | 0.125 | 0     | 0     | 0     | 0     | 0     | 0     | 0     |
| commonality | 0  | 0     | 0     | 0.143 | 0     | 0     | 0     | 0     | 0     | 0     |
| internet | 0     | 0     | 0     | 0.143 | 0     | 0     | 0     | 0     | 0     | 0     |
| group    | 0     | 0     | 0     | 0     | 0.080 | 0     | 0.036 | 0     | 0.066 | 0.057 |
| contact  | 0     | 0     | 0     | 0     | 0     | 0.125 | 0     | 0     | 0     | 0     |
| experience | 0   | 0     | 0     | 0     | 0.2   | 0     | 0     | 0     | 0     | 0     |
| purpose  | 0     | 0     | 0     | 0     | 0.2   | 0     | 0     | 0     | 0     | 0     |
| prospect | 0     | 0     | 0     | 0     | 0     | 0.125 | 0     | 0     | 0     | 0     |
| support  | 0     | 0     | 0     | 0     | 0     | 0.125 | 0     | 0     | 0     | 0     |
| computer | 0     | 0     | 0     | 0     | 0     | 0     | 0.072 | 0.199 | 0.066 | 0.114 |
| device   | 0     | 0     | 0     | 0     | 0     | 0     | 0.091 | 0     | 0     | 0     |
| channel  | 0     | 0     | 0     | 0     | 0     | 0     | 0.091 | 0     | 0     | 0     |
| communication | 0 | 0    | 0     | 0     | 0     | 0     | 0.091 | 0     | 0     | 0     |
| user     | 0     | 0     | 0     | 0     | 0     | 0     | 0.182 | 0     | 0     | 0     |
| resource | 0     | 0     | 0     | 0     | 0     | 0     | 0.091 | 0     | 0     | 0     |
| data     | 0     | 0     | 0     | 0     | 0     | 0     | 0.064 | 0     | 0     | 0.100 |
| server   | 0     | 0     | 0     | 0     | 0     | 0     | 0     | 0.25  | 0     | 0     |
| cable    | 0     | 0     | 0     | 0     | 0     | 0     | 0     | 0     | 0.167 | 0     |
| signal   | 0     | 0     | 0     | 0     | 0     | 0     | 0     | 0     | 0.167 | 0     |
| protocol | 0     | 0     | 0     | 0     | 0     | 0     | 0     | 0     | 0.167 | 0     |

Table 5: TF * IDF.

**FP-Growth and FCM Approach:**
minimum support = 3

|  | Frequent Sets | Frequency |
|---|---|---|
| Set 1 | D2, D4, D6 | 3 |
| Set 2 | D7, D9, D10 | 3 |

Table 6: Frequent sets obtained from FP-growth.

|  | Cluster Center |
|---|---|
| Center for cluster 1 | ( D2 + D4 + D6 )/ 3 = <br> ( 0.0349 0 0.0346 0.0218 0.0667 0.0815 0.0815 0.0212 0 0 0 0.0476 0.0476 0 0.0417 0 0 0.0417 0.0417 0 0 0 0 0 0 0 0 0 0 0 ) |
| Center for cluster 2 | ( D7 + D9 + D10 )/ 3 = <br> ( 0 0 0 0 0 0 0 0.0354 0 0 0 0 0 0 0.0531 0 0 0 0 0 0.0841 0.0303 0.0303 0.0303 0.0606 0.0303 0.0545 0 0.0556 0.0556 0.0556) |

Table 7: Initial Cluster Centers.

m = 2:

| Cluster | D1 | D2 | D3 | D4 | D5 | D6 | D7 | D8 | D9 | D10 |
|---|---|---|---|---|---|---|---|---|---|---|
| cluster1 | .5411 | .6035 | .5139 | .5832 | .5050 | .5884 | .4187 | .4323 | .4337 | .3431 |
| cluster2 | .4589 | .3965 | .4861 | .4168 | .4950 | .4116 | .5813 | .5677 | .5663 | .6569 |

m = 1.5

| Cluster | D1 | D2 | D3 | D4 | D5 | D6 | D7 | D8 | D9 | D10 |
|---|---|---|---|---|---|---|---|---|---|---|
| cluster 1 | 0.6116 | 0.8114 | 0.5366 | 0.7758 | 0.5078 | 0.7828 | 0.2325 | 0.3142 | 0.2552 | 0.1365 |
| cluster 2 | 0.3884 | 0.1886 | 0.4634 | 0.2242 | 0.4922 | 0.2172 | 0.7675 | 0.6858 | 0.7448 | 0.8635 |

Table 8: Results of Fuzzy C-Means.

C1: (0.0432 0.0314 0.0333 0.0382 0.0384
0.0454 0.0454 0.0194 0.0258 0.0129
0.0129 0.0256 0.0256 0.0116 0.0227
0.0190 0.0190 0.0227 0.0227 0.0151
0.0027 0.0027 0.0027 0.0054 0.0027
0.0032 0.0116 0.0056 0.0056 0.0056 )

C2: (0.0157 0.0156 0.0122 0.0192 0.0042
0.0060 0.0060 0.0310 0.0203 0.0102
0.0102 0.0039 0.0039 0.0361 0.0033
0.0178 0.0178 0.0033 0.0033 0.0762
0.0158 0.0158 0.0158 0.0315 0.0158
0.0317 0.0366 0.0276 0.0276 0.0276 )

Table 9: Cluster Centroids.

| Cluster 1 | D1, D2, D3, D4, D5, D6 |
|---|---|
| Cluster 2 | D7, D8, D9, D10 |

Table 10: Final clusters.

**FP Growth and K-means Approach:**
minimum support = 3

| Set 1 | D2, D4, D6 | 3 |
|---|---|---|
| Set 2 | D7, D9, D10 | 3 |

Frequent sets obtained from FP-growth.

Initial clusters for iteration 1

|  | Cluster Center |
|---|---|
| Center for Cluster 1 | ( D2 + D4 + D6 )/ 3 = ( 0.0349 0 0 0.0346 0.0218 0.0667 0.0815 0.0815 0.0212 0 0 0 0.0476 0.0476 0 0.0417 0 0 0.0417 0.0417 0 0 0 0 0 0 0 0 0 0 ) |
| Center for Cluster 2 | ( D7 + D9 + D10 )/ 3 = ( 0 0 0 0 0 0 0 0.0354 0 0 0 0 0 0.0531 0 0 0 0 0 0.0841 0.0303 0.0303 0.0303 0.0606 0.0303 0.0545 0 0.0556 0.0556 0.0556) |

Distance from the initial clusters for iteration 1

|  | D1 | D2 | D3 | D4 | D5 | D6 | D7 | D8 | D9 | D10 |
|---|---|---|---|---|---|---|---|---|---|---|
| Cluster 1 | 0.334754 | 0.185787 | 0.350747 | 0.173635 | 0.350191 | 0.183109 | 0.327575 | 0.364496 | 0.35048 | 0.242285 |
| Cluster 2 | 0.361395 | 0.325223 | 0.358166 | 0.287677 | 0.348769 | 0.300186 | 0.199287 | 0.313358 | 0.219351 | 0.143418 |

Clusters after iteration 1:
Cluster1: D1, D2, D3, D4, D6
Cluster2: D5, D7, D8, D9, D10

New centroids for iteration 2

|  | Cluster Center |
|---|---|
| Center for Cluster 1 | ( D1 + D2 + D3+ D4 + D6)/ 5 = (0.0602,0.05,0.0374,0.0392,0.04,0.049,0.049,0.0204,0.05,0.025,0.025,0.0286,0.0286,0,0.025,0,0,0.025   0.025,0,0,0,0,0,0,0,0,0,0,0) |
| Center for Cluster 2 | ( D5 + D7 + D9 + D9 +D10 )/ 5 = (0 0 0.0088 0.021    0 0 0 0.029 0 0 0 0 0.0478 0 0.04 0.04 0 0 0.0902 0.0182 0.0182 0.0182    0.0364 0.0182 0.0328 0.05 0.0334 0.0334 0.0334) |

Distance from the centroids for iteration 2

|  | D1 | D2 | D3 | D4 | D5 | D6 | D7 | D8 | D9 | D10 |
|---|---|---|---|---|---|---|---|---|---|---|
| Cluster1 | 0.260439 | 0.343409 | 0.213627 | 0.314308 | 0.2727763 | 0.349397 | 0.207361 | 0.276669 | 0.333131 | 0.281214 |
| Cluster2 | 0.343409 | 0.213627 | 0.314308 | 0.2727763 | 0.349397 | 0.207361 | 0.276669 | 0.333131 | 0.281214 | 0.210661 |

Clusters after iteration 2:
Cluster1: D1, D2, D3, D4, D6
Cluster2: D5, D7, D8, D9, D10

The final clusters are
Cluster1: (D1, D2, D3, D4, D6)
Cluster2: (D5, D7, D8, D9, D10)

**FP Growth and Cosine_Similarity Approach:**
minimum support = 3

| Set 1 | D2, D4, D6 | 3 |
|---|---|---|
| Set 2 | D7, D9, D10 | 3 |

Frequent sets obtained from FP-growth.

The uncluster documents are D1, D3, D5, D8.

The final clusters are
Cluster1 :(D1, D2, D3, D4, D6)
Cluster2: (D5, D7, D8, D9, D10)